\documentclass{article}

\usepackage{PRIMEarxiv}

\usepackage[utf8]{inputenc} 
\usepackage[T1]{fontenc}    
\usepackage{hyperref}       
\usepackage{url}            
\usepackage{booktabs}       
\usepackage{amsfonts}       
\usepackage{nicefrac}       
\usepackage{microtype}      
\usepackage{lipsum}
\usepackage{fancyhdr}       
\usepackage{graphicx}       
\graphicspath{{media/}}     
\usepackage{comment}
\usepackage{titlesec}
\titleformat{name=\section,numberless}{\normalfont\bfseries\small}{}{0pt}{}
\titlespacing{\section}{0pt}{3ex}{1ex}
\usepackage[round]{natbib}

\usepackage{amsmath}

\title{Entropy in Sicence of Science
\thanks{\textit{\underline{Citation}}: 
\textbf{Authors. Title. Pages.... DOI:000000/11111.}} 
}

\author{
  Yujie Shi \\
  School of Information Management \\
  Nanjing University \\
  Nanjing, China\\
  \texttt{yujieshi@smail.nju.edu.cn} \\
     \And
  Alex Jie Yang \\
  School of Information Management \\
  Nanjing University \\
  Nanjing, China\\
  \texttt{alexjieyang@outlook.com} \\
   \And
  Sanhong Deng \\
  School of Information Management \\
  Nanjing University \\
  Nanjing, China\\
  \texttt{sanhong@smail.nju.edu.cn} \\
}

\begin{document}
\maketitle

\begin{abstract}
This study investigates entropy's potential for analyzing scientific research patterns across disciplines. Originating from thermodynamics, entropy now measures uncertainty and diversity in information systems. We examine Shannon Entropy, Entropy Weight Method, Maximum Entropy Principle and structural entropy applications in scientific collaboration, knowledge networks, and research evaluation. Through publication analysis and collaboration network studies, entropy-based approaches demonstrate effectiveness in mapping interdisciplinary knowledge integration, with higher entropy values correlating to increased knowledge diversity in citation networks. Structural entropy analysis reveals dynamic collaboration patterns affecting research productivity. Results indicate entropy metrics offer objective tools for assessing research quality, optimizing team structures, and informing science policy decisions. These quantitative methods enable systematic tracking of knowledge evolution and resource allocation efficiency, providing actionable insights for researchers and policymakers managing complex scientific ecosystems
\end{abstract}


\keywords{Entropy applications\and Shannon Entropy\and Entropy Weight Method\and Maximum Entropy Principle\and Science of Science\and Network Analysis\and Academic Evaluation}

\section{Introduction}
Entropy first appeared in thermodynamics, thanks to Clausius and his work on the second law. Then, in 1948, Claude E. Shannon came up with his "statistical theory of communication" \citep{shannon1948mathematical}. He took the idea of entropy and made it useful for many different fields. Since Shannon’s breakthrough, information entropy has become really important in areas like communication systems, machine learning, biology, and finance.

Science of Science, or scientometrics, started in the mid-20th century. It aims to study scientific practices, funding, and their effects using numbers and data. A key book in this area is Derek J. de Solla Price’s \textit{Little Science, Big Science} \citep{price1963little} from 1963, which set the stage for future work. Around the same time, Eugene Garfield made big strides with citation analysis, especially through the Science Citation Index \citep{leydesdorff2001challenge}. This field looks at research networks, how papers cite each other, and how new ideas spread. Lately, it’s grown a lot by using big data and network science.

H. Grupp’s paper, "The concept of entropy in scientometrics and innovation research" \citep{grupp1990concept}, is an important early example of using information entropy in scientometrics. In the study, Grupp uses information entropy to measure how much research institutions are involved in R\&D and to look at how broad and deep national technology strategies are. By using data from bibliometrics and patents, Grupp shows how useful information entropy can be for analyzing science. Since then, people have kept using information entropy in Science of Science, which has helped a lot in understanding the complex parts of scientific systems \citep{basurto2018entropy}.

This paper aims to give a detailed look at today’s information entropy theories and how they’re used in Science of Science. Here’s how the rest of the paper is set up: Section 2 follows the history of entropy from thermodynamics to information theory, talking about Clausius entropy, Boltzmann-Gibbs entropy, and Shannon entropy. Section 3 reviews Science of Science, including what it is and what people are studying in it now. Section 4 looks at other work related to using entropy. Section 5 explains Shannon information entropy and some related ideas like relative entropy, cross entropy, and mutual information, and how they’re used in Science of Science. Section 6 talks about other kinds of entropy that are often used in Science of Science, like Renyi entropy, the entropy weight method, the maximum entropy principle, and structural entropy. Section 7 shows how entropy is used in practice in Science of Science, especially for measuring interdisciplinary research, mapping knowledge, and evaluating academic work.

\begin{figure}[!htp]%
\centering
\includegraphics[width=1\textwidth]{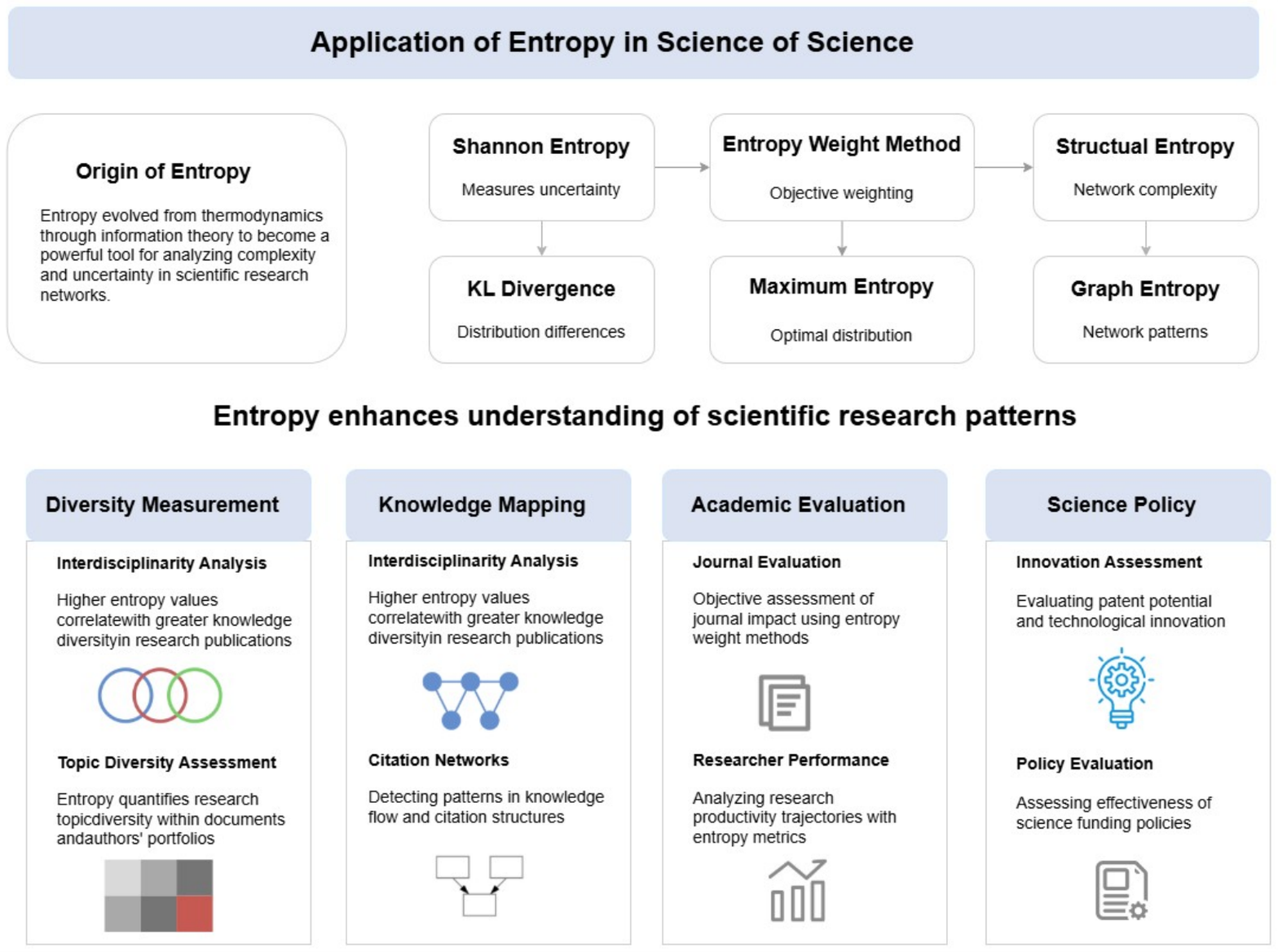}
\caption{Application of Entropy in Science of Science}\label{fig1}
\end{figure}

\section{From Thermodynamics to Information Entropy}
\subsection{Clausius Entropy: The Macroscopic Perspective of Thermodynamics}
Entropy came from studies in the 1800s about how energy changes form. While looking at how efficient heat engines are and why heat flows in one direction, Clausius came up with entropy ($S$) as a way to describe the state of a system. The equation for it is:
\begin{equation}
dS = \frac{\delta Q}{T} + \delta S_{\text{gen}}
\end{equation}
Here, $\delta Q$ is the heat the system takes in, $T$ is the temperature, and $\delta S_{\text{gen}}$ is the extra entropy from processes that can’t be reversed. Clausius pointed out that entropy doesn’t go backwards: in a system that’s closed off, if something happens that can’t be undone ($\delta S_{\text{gen}} > 0$), the entropy goes up. But if the process can be reversed ($\delta S_{\text{gen}} = 0$), entropy only changes because of heat moving in or out. This idea is the math behind the Second Law of Thermodynamics, which says that in a closed system, entropy never gets smaller \citep{cover2012elements}.

Clausius’s entropy linked big-scale events with how energy moves and stays the same, but it didn’t explain what was happening at the tiny, particle level. That left room for later work in statistical mechanics \citep{greven2014entropy}.

\subsection{Boltzmann-Gibbs Entropy: The Microscopic Interpretation from Statistical Mechanics}
Toward the end of the 1800s, Boltzmann and Gibbs looked at how molecules move and saw entropy as a way to measure the chances of different tiny states. Boltzmann said that for an ideal gas, entropy is related to the logarithm of how many microstates there are, written as $W$:
\begin{equation}
S = k \log W \quad \text{(assuming each state is equally likely)}
\end{equation}
Gibbs took this further for cases where states aren’t equally likely, coming up with a general formula for statistical entropy:
\begin{equation}
S = -k \sum_{i=1}^{n} p_i \log p_i
\end{equation}
In this equation, $p_i$ is the chance of the $i$-th microstate, and $k$ is the Boltzmann constant. This shows that entropy is basically the average uncertainty about the system’s microstates. If all microstates are equally likely, it goes back to Boltzmann’s formula. But if some states are more likely than others, entropy is less. This way of thinking about entropy connected the idea that tiny events can go backwards with the fact that big-scale events don’t, showing how entropy links big thermodynamic numbers to tiny probability ideas. This set the stage for using entropy in many different fields \citep{wehrl1978general}.

\subsection{Shannon Entropy: A Measure of Uncertainty in Information Theory}
In the middle of the 20th century, Shannon gave entropy a new meaning in information theory, making it a way to measure how uncertain information is. For a random variable $X$ that can take different values with probabilities $p(x)$, Shannon entropy is:
\begin{equation}
H(X) = -\sum_{i=1}^n p(x_i) \log_2 p(x_i)
\end{equation}
This looks just like the Boltzmann-Gibbs entropy, but it’s about something different: information entropy tells us the average amount of information in a signal, which is key for figuring out how much we can compress data. For instance, if a signal’s probabilities aren’t all the same, we can use tricks like Huffman coding to make the average number of bits smaller, and the best we can do is set by $H(X)$. Shannon called it "entropy" because von Neumann suggested it, since both ideas deal with uncertainty and have similar math \citep{shannon1948mathematical}.

Bringing in information entropy went beyond just physical systems. Because it doesn’t depend on specific units and works with any probability distribution, it’s been used in lots of areas like communications, biology, and computer science, becoming a big deal in the information age \citep{cover2012elements}.

\section{A review of Science of Science and its Research Fields}
Science of Science (SciSci) is a field that crosses many disciplines. It uses big data and computer tools to find patterns in how science works. It looks at how scientific knowledge is made and shared, how scientists act, and how the scientific community changes over time. Thanks to databases like Web of Science, Scopus, and PubMed, researchers can dig into huge amounts of publication data, which has really pushed the field forward. SciSci has its roots in 20th-century philosophy and sociology of science, with important ideas from Kuhn and Merton \citep{kuhn1962structure, merton1973sociology}. It’s important because it gives real evidence and theories to help science move forward, make research management better, and guide policy decisions.

\subsection{Production and Dissemination of Scientific Knowledge}
In SciSci, making and sharing scientific knowledge are key topics. Scientific discoveries usually come in two types: small steps that improve what’s already known, and big leaps that change everything. Often, the most influential science comes from mixing standard and new ideas, as Uzzi et al. showed. They found that breakthroughs happen when deep knowledge and fresh perspectives meet in citation networks \citep{uzzi2013atypical}.

Sharing knowledge happens through citation networks and collaboration networks. Citation networks show how knowledge moves from one paper to another; for example, Chen et al. used citation analysis to find patterns in how discoveries are made \citep{chen2009towards}. Collaboration networks show how scientists work together, and when people from different fields team up, it can really help combine knowledge. Publishing and peer review also affect how knowledge spreads, but there are problems like the "file drawer problem," where negative results don’t get published as much, which can make the scientific record less accurate.

\subsection{Scientists’ Careers and Behaviors}
SciSci looks at how scientists’ careers develop and how they behave, which are important areas of study. Moving from being a graduate student to a senior researcher depends on things like education, what they research, and who they work with. When choosing what to study, scientists have to decide between sticking with what’s known, which is safer, or trying something new, which could lead to big discoveries \citep{uzzi2013atypical}.

Working together and competing are big parts of how science moves forward. Wu et al. found that small teams often come up with ideas that shake things up, while bigger teams focus on hot topics and make a splash, but it doesn’t last long \citep{wu2019large}. To measure how productive and influential scientists are, people often use the h-index, but it needs to be tweaked to fit different fields and individual work.

\subsection{Structure and Dynamics of the Scientific Community}
The scientific community is like a big, complicated network, and we can study it using social network tools. Newman used collaboration networks to show how scientists are organized \citep{newman2001structure}. Citation networks help us see how different fields are connected and how knowledge relies on other knowledge.

How scientific fields change over time involves new areas popping up and blending together, like how biomedical sciences have grown by mixing different disciplines. The rules and culture of science are really important for how the community grows; for example, open science has made people share data more, but it also brings up issues about being open and keeping things private.

\subsection{Science Policy and Research Management}
What we learn from SciSci helps shape science policy and how research is managed. How research money is given out affects what science produces, as Azoulay et al. showed that the way funding is set up can really change how much innovation happens \citep{azoulay2011incentives}. When evaluating research, people often use numbers like impact factors and how many times something is cited, but these don’t always show the whole picture, like when people cite others just to be polite. Science policy can guide research by focusing on projects that are risky but could pay off big.

\subsection{Methodology}
SciSci uses both numbers and stories to understand science. Scientometrics, which includes bibliometrics and network analysis, is a key part of it; Garfield was the first to use citation analysis to evaluate scientific papers \citep{garfield1972citation}. On the other hand, qualitative methods like case studies give us a deeper look at how science works.

The data comes from places like Web of Science, PubMed, and patent databases (like USPTO), which have lots of information on publications, funding, and patents. But there are problems, like most of the data being in English and not everyone having the same access to it.

\section{Related Work}
There are many great reviews about entropy that look at its history or how it’s used in different areas.

Li et al. did a study on the journal Entropy, showing how research topics have changed over 20 years, with things like graph entropy, permutation entropy, and pseudo-additive entropy becoming hot topics \citep{li2019twenty}. Sepúlveda-Fontaine et al. organized entropy theories from the classics (like Boltzmann-Gibbs and Shannon entropy) to newer ones (like approximate entropy, cross-entropy, Rényi entropy, and Tsallis entropy), and they really focused on how these are used in machine learning and data analysis \citep{sepulveda2024applications}. Ribeiro et al. \citep{ribeiro2021entropy} were among the first to use entropy in analyzing time series, while Sabirov et al. \citep{sabirov2021information} and Ma et al. \citep{ma2018shannon} set up ways to use entropy in chemistry and physics, specifically in making chemicals and studying heavy-ion collisions. Amigó et al. \citep{amigo2015entropy} built a math foundation that shows how entropy can be used in many different sciences. Katok’s study \citep{katok2007fifty} from 1958 to 2007 revealed how the main ideas about entropy have evolved, pointing out key theorems that changed how we think about it. Together, these works create a full picture that connects basic entropy ideas with complex systems.

However no research has really looked at how entropy is used in Science of Science. We suggest reading these reviews along with our paper because Science of Science pulls together methods from many fields. Even though they focus on different things, these reviews add to what we’re saying.

\section{Shannon Information Entropy and Relevant Concepts}

\subsection{Shannon Information Entropy}

Information entropy, introduced by Claude Shannon in 1948~\citep{shannon1948mathematical}, lies at the heart of information theory. It measures how uncertain or random information is. This idea builds on probability theory and shows the unpredictability of a random variable. If all events have an equal chance of happening, entropy increases, reflecting higher uncertainty. But if some events are far more likely, entropy decreases.

The formula for information entropy \( H(X) \) of a random variable \( X \) is:

\begin{equation}
H(X) = -\sum_{i=1}^n p(x_i) \log_2 p(x_i)
\end{equation}

where:
\begin{itemize}
    \item \( p(x_i) \) is the probability that \( X \) equals \( x_i \),
    \item the logarithm uses base 2, so entropy is expressed in bits.
\end{itemize}

In essence, this formula captures the average information gained when an event occurs. Events that rarely happen carry more information because they surprise us, adding more to the entropy.

\subsection{Relative Entropy/KL Divergence}

Relative entropy, often called Kullback–Leibler (KL) divergence, measures how much two probability distributions \( P \) and \( Q \) differ~\citep{kullback1951information}. It is given by:

\begin{equation}
D_{KL}(P||Q) = \sum_{i=1}^n p_i \log \frac{p_i}{q_i}
\end{equation}

Its key properties are:
\begin{enumerate}
    \item It is never negative: \( D_{KL}(P||Q) \geq 0 \). It equals zero only when \( P \) and \( Q \) are the same. This follows from Gibbs' inequality, which says that for distributions where \( \sum_{i=1}^n p_i = \sum_{i=1}^n q_i = 1 \) and \( p_i, q_i \in (0,1) \), \( -\sum_{i=1}^n p_i \log p_i \leq -\sum_{i=1}^n p_i \log q_i \). Equality holds only if \( p_i = q_i \) for all \( i \).
    \item It lacks symmetry: \( D_{KL}(P||Q) \neq D_{KL}(Q||P) \). This means the direction of comparison matters.
\end{enumerate}

To clarify:
\begin{itemize}
    \item \( D_{KL}(P||Q) \) shows how far \( Q \) strays from the true distribution \( P \),
    \item \( D_{KL}(Q||P) \) does the reverse, using \( Q \) as the reference.
\end{itemize}

\subsection{Cross Entropy}

Cross entropy \( H(P, Q) \) gauges how well an estimated distribution \( Q \) matches the true distribution \( P \)~\citep{cover2006elements}. It is defined as:

\begin{equation}
H(P, Q) = -\sum_{i=1}^{n} p_i \log q_i
\end{equation}

where:
\begin{itemize}
    \item \( p_i \) comes from the true distribution,
    \item \( q_i \) comes from the estimated one.
\end{itemize}

Cross entropy combines the entropy of \( P \) with the KL divergence between \( P \) and \( Q \):

\begin{equation}
H(P, Q) = H(P) + D_{KL}(P || Q)
\end{equation}

This equation reflects both the natural uncertainty in \( P \) and the additional uncertainty from using \( Q \).

\subsection{Joint Information Entropy}

Joint entropy captures the uncertainty tied to two random variables considered together. For variables \( X \) and \( Y \) with a joint probability \( p(x,y) \), it is:

\begin{equation}
H(X,Y) = -\sum_{x \in \mathcal{X}} \sum_{y \in \mathcal{Y}} p(x,y) \log p(x,y)
\end{equation}

Here, \( \mathcal{X} \) and \( \mathcal{Y} \) are the sets of possible values for \( X \) and \( Y \).

This measure shows the average information needed to describe both variables at once~\citep{shannon1948mathematical}. It finds use in several areas:
\begin{itemize}
    \item In image processing, it reveals how pixels relate to each other.
    \item In natural language processing, it explores links between words.
    \item In optimization, it assesses connections between variables
\end{itemize}

\subsection{Mutual Information Entropy}

Mutual information (MI) reveals how much one random variable tells us about another. It quantifies the shared information between \( X \) and \( Y \). One way to see it is as the drop in uncertainty about one variable when the other is known:

\begin{equation}
I(X;Y) = H(X) - H(X|Y) = H(Y) - H(Y|X)
\end{equation}

where:
\begin{itemize}
    \item \( H(X) \) is the entropy of \( X \),
    \item \( H(X|Y) \) is the entropy of \( X \) given \( Y \).
\end{itemize}

Another form is:

\begin{equation}
I(X;Y) = \sum_{x \in \mathcal{X}} \sum_{y \in \mathcal{Y}} p(x,y) \log \frac{p(x,y)}{p(x)p(y)}
\end{equation}

Its properties include:
\begin{enumerate}
    \item It is non-negative: \( I(X;Y) \geq 0 \). It is zero only if \( X \) and \( Y \) are independent.
    \item It is symmetric: \( I(X;Y) = I(Y;X) \).
    \item It measures the information one variable provides about the other.
\end{enumerate}

\subsection{Conditional Entropy}

Conditional entropy tracks the uncertainty about \( X \) when \( Y \) is known. For variables \( X \) and \( Y \), it is:

\begin{equation}
H(X|Y) = -\sum_{x \in \mathcal{X}} \sum_{y \in \mathcal{Y}} p(x,y) \log p(x|y)
\end{equation}

where \( p(x|y) \) is the probability of \( X = x \) given \( Y = y \).

It can also be written as:

\begin{equation}
H(X|Y) = H(X,Y) - H(Y)
\end{equation}

This shows it as the total uncertainty of both variables minus the uncertainty of \( Y \) alone.

\subsection{Renyi Entropy}

Renyi entropy, proposed by Alfred Renyi in 1961~\citep{renyi1961measures}, extends Shannon entropy. It measures uncertainty with a parameter \( q \). For a distribution \( P = (p_1, p_2, \ldots, p_n) \), it is:

\begin{equation}
H_q(P) = \frac{1}{1 - q} \log \left( \sum_{i=1}^n p_i^q \right)
\end{equation}

The value of \( q \) affects it:
\begin{itemize}
    \item At \( q = 0 \), it becomes \( H_0(P) = \log n \), where \( n \) is the number of outcomes.
    \item At \( q = 1 \), it matches Shannon entropy.
    \item For other \( q \), it offers varied ways to assess uncertainty, useful in statistics and quantum theory.
\end{itemize}

\subsection{Application of Information Entropy in Science of Science}

Information entropy has proven valuable in studying science itself, offering ways to analyze communication, collaboration, and knowledge patterns. Since Shannon’s work in 1948, it has helped evaluate research quality and impact. For instance, it measures interdisciplinarity~\citep{mutz2022diversity, leydesdorff2011indicators, wagner2011approaches, kim2024identifying, park2023papers, kim2024persistent}. Higher entropy signals more field diversity. The entropy weight method aids journal evaluation~\citep{zhao2023research, duan2015journals, wang2024research, xu2023multidimensional}, reducing biases like the Matthew effect in citations.

Entropy also supports keyword analysis~\citep{noh2015keyword, zhang2019research, lou2021semantic}, picking out key terms in papers. In citation studies, it evaluates research efficiency~\citep{prathap2011comments}, model predictions~\citep{madani2018keyword}, and topic focus~\citep{daud2019correlational}. Mutual information and conditional entropy shed light on collaboration networks~\citep{li2024academic, riahinia2022synergistic}, showing how knowledge spreads. The maximum entropy principle applies to bibliometric analysis~\citep{lafouge2001links} and author contributions~\citep{gerchak2020inferring}.

These uses highlight how entropy provides clear, quantitative insights into science, beyond just citation counts, aiding policy and research management.

\section{Other Common Used Entropy related Theories in Science of Science}

\subsection{Entropy Weight Method (EWM)}

The Entropy Weight Method, or EWM, is a way to assign weights to different factors in a decision-making process. It draws on the concept of information entropy from information theory. People often use this method for tasks like evaluating multiple criteria, analyzing decisions, and improving systems. The key idea is to measure each factor’s importance by calculating its information entropy. This shows how much each factor matters in the final decision.

At its core, EWM relies on information entropy to check how spread out the data is for each indicator. Information entropy, a basic idea in information theory, measures the amount of uncertainty or disorder in a system. When evaluating multiple criteria, it helps reveal the variation in indicator data. Here’s how it breaks down:
\begin{itemize}
    \item \textbf{Information Entropy:} High entropy means the indicator’s data is widely spread, carrying more information. Low entropy suggests the data is more uniform, with less information.
    \item \textbf{Weight Calculation:} We calculate weights using the entropy values. Indicators with lower entropy get higher weights, while those with higher entropy get lower weights.
\end{itemize}

\section*{Computational Steps}

To apply the Entropy Weight Method, we go through these steps:

\begin{itemize}
   \item \textbf{Standardization}: Indicators often have different units or scales, so we standardize the data to make them comparable. This removes the effects of different dimensions. Two common approaches are:
   \begin{itemize}
       \item Min-max normalization, which adjusts the data to fit between 0 and 1:
       \begin{equation}
       Z_{ij} = \frac{x_{ij}-\min(X_j)}{\max(X_j)-\min(X_j)}
       \end{equation}

       \item Z-score standardization, which uses the mean and standard deviation:
        \begin{equation}
       Z_{ij} = \frac{X_{ij}-\bar{X_j}}{s_j}
       \end{equation}
\end{itemize}
\end{itemize}

\begin{itemize}
   \item \textbf{Calculation of Sample Proportions}: For each indicator $j$, we figure out the proportion of the $i$-th sample:
          \begin{equation}
   P_{ij} = \frac{Z_{ij}}{\sum_{i=1}^{n} Z_{ij}}
       \end{equation}

   Here, $n$ is the total number of samples.
\end{itemize}

\begin{itemize}
   \item \textbf{Calculation of Information Entropy}: Next, we use those proportions to find the entropy for each indicator $j$:

          \begin{equation}
   H(X_j) = -\sum_{i=1}^{n} P_{ij} \ln(P_{ij})
       \end{equation}

   This $H(X_j)$ shows the entropy of the $j$-th indicator.
\end{itemize}

\begin{itemize}
   \item \textbf{Weight Calculation}: Finally, we determine each indicator’s weight based on its entropy:
             \begin{equation}
   w_j = \frac{1-H(X_j)}{\sum_{j=1}^{m} (1-H(X_j))}
       \end{equation}

   Where:
   \begin{itemize}
       \item $w_j$ is the weight of the $j$-th indicator
       \item $m$ is the total number of indicators
       \item $H(X_j)$ comes from the previous step
   \end{itemize}
\end{itemize}

\subsubsection*{Applications of Entropy Weight Method in Science of Science}

The Entropy Weight Method (EWM) finds use across many areas in the science of science. Here are some notable examples:

\textbf{Academic Journal Evaluation}  
Researchers have applied EWM to assess journals from different angles, such as their impact, reach, and innovation. Studies like those by Zhao et al. \citep{zhao2023research}, Duan et al. \citep{duan2015journals}, Wang et al. \citep{wang2024research}, and Xu et al. \citep{xu2023multidimensional} show it reduces subjective bias and the Matthew effect—where well-known journals gain more attention unfairly. This leads to fairer evaluations across various fields.

\textbf{Research Performance Assessment}  
EWM helps measure research performance at different levels. B\u{a}din et al. \citep{buadin2018reflecting} and \c{S}erban et al. \citep{cserban2017appraisal} used it to build frameworks for evaluating research capabilities of institutions and countries. Liu et al. \citep{liu2018evaluate} ranked research teams by their academic influence, while Sheng et al. \citep{shengresearch} created models to assess individual researchers more objectively.

\textbf{Technological Innovation Assessment}  
Zhang et al. \citep{zhang2017entropy} were early adopters, using EWM to evaluate patents for their innovation potential. Their approach spots patents that excel in specific areas, avoiding the averaging-out problem of older methods. Paired with techniques like collaborative filtering, it picks out promising patents from large datasets.

\textbf{Other Applications}  
EWM also applies to other areas, including:
\begin{itemize}
    \item Checking how well companies adapt to innovation shifts \citep{xia2023research}
    \item Evaluating policy effectiveness \citep{cui2022policy}
    \item Measuring talent ecosystem growth \citep{liang2024research}
    \item Spotting emerging technologies \citep{zhang2021bi}
\end{itemize}

The strength of EWM lies in its use of data patterns to set weights objectively. This avoids the pitfalls of subjective methods, making it great for tackling complex evaluation tasks in scientometric research.

\subsection{The Maximum Entropy Principle (MEP)}

The Maximum Entropy Principle (MEP) is a key idea in information theory and statistics. It guides us in choosing the best probability distribution when we only have partial information about a system \citep{jaynes1957information}. The principle says to pick the distribution with the highest entropy that fits the known constraints. Entropy reflects uncertainty—higher entropy means more uncertainty and more information potential in the system.

Put simply, when we lack full details, MEP suggests assuming the system is as uniform as possible. This keeps us from jumping to conclusions without evidence.

\subsubsection*{Applications of MEP in Science of Science}

The Maximum Entropy Principle has several uses in scientometric research:
\begin{itemize}
    \item Lafouge and Michel \citep{lafouge2001links} used it to study bibliometric distributions, tracking entropy changes to explore links between author productivity, keywords, and information entropy.
    \item Susan and Keshari \citep{susan2019finding} applied a Maximum Entropy Partitioning method to pinpoint key terms, blending term frequency and distinctiveness. Their tests showed better classification with fewer features.
    \item Dainelli and Saracco \citep{dainelli2023bibliometric} used MEP to build a model for spotting patterns in citation networks, making bibliometric analysis clearer.
    \item Gerchak \citep{gerchak2020inferring} used entropy maximization to estimate author contributions in papers, analyzing name order deviations for a quantitative approach.
    \item Li et al. \citep{yong2013advisor} developed an MEP-based method to identify advisor-advisee ties in co-authorship networks, hitting over 95\% accuracy on DBLP data.
\end{itemize}

\subsection{Structural Entropy}

Structural entropy measures how complex a network’s structure is. Li and Pan introduced it in 2016 \citep{li2016structural}. Unlike typical entropy, which focuses on randomness, this looks at patterns and regularities in networks.

The concept treats network nodes as points in a high-dimensional space. We explore the network with random walks and encode node positions. The $K$-dimensional structural entropy, $HK(G)$, is the smallest number of bits needed to encode the $K$-dimensional coordinates of reachable nodes. As $K$ grows, the captured structure gets more intricate.

\subsubsection*{Applications of Structural Entropy in Science of Science}

Structural entropy has useful applications in science of science:
\begin{itemize}
    \item Xu et al. \citep{xu2022methodology} combined it with link prediction to find breakthrough research topics, tracking structural entropy changes over time.
    \item Liu and Gao \citep{liu2023structure} used it to rank node importance, measuring entropy shifts after removing nodes to assess their role in connectivity.
    \item Cunha et al. \citep{cunha2020shannon} applied it to semantic networks of paper titles, tracing new ideas and research trends.
\end{itemize}

\subsection{Graph Entropy}

Graph entropy uses information theory to measure a graph’s structural complexity. Rashevsky \citep{rashevsky1955life} and Trucco \citep{trucco1956a} first proposed it in the 1950s to capture a graph’s information content.

Its basic formula builds on Shannon’s entropy, splitting the graph into equivalent parts:
\begin{equation}
I(G,a) = \sum_{i=1}^{k} \frac{|X_i|}{|X|} \log \frac{|X_i|}{|X|}
\end{equation}
Here, $|X_i|$ is the size of the $i$-th group, and $|X|$ is the total number of invariants.

Later, researchers expanded on this:
\begin{itemize}
    \item Mowshowitz \citep{mowshowitz1968entropya,mowshowitz1968entropyb,mowshowitz1968entropyc,mowshowitz1968entropyd} explored entropy using symmetries and colorings.
    \item Körner \citep{korner1973coding} tied it to coding theory.
    \item Bonchev and Trinajstić \citep{bonchev1977information} used distance matrices in the 1970s-80s:
    \begin{equation}
    I_D(G) = \sum_{i=1}^{\rho(G)} \frac{2k_i}{|V|^2} \log \frac{2k_i}{|V|^2}
    \end{equation}
    \item Hosoya \citep{hosoya1971topological} based it on polynomial coefficients.
\end{itemize}
More recently, Dehmer \citep{dehmer2008a,dehmer2008structural} introduced a flexible measure using vertex probabilities:
\begin{equation}
I_f(G) := -\sum_{i=1}^{|V|} \frac{f(v_i)}{\sum_{j=1}^{|V|}f(v_j)} \log \left(\frac{f(v_i)}{\sum_{j=1}^{|V|}f(v_j)}\right)
\end{equation}

\subsubsection*{Applications of Graph Entropy in Science of Science}

Graph entropy plays a big role in science of science, especially for networks like collaborations or citations:
\begin{itemize}
    \item In information theory, it optimizes data flow in networks, building on Cover et al.’s work \citep{cover2006elements}.
    \item Dehmer \citep{dehmer2008information} offered a framework to measure network information capacity.
    \item In bioinformatics, Dehmer and Mowshowitz \citep{dehmer2011a} used it to study gene and protein networks.
    \item Cao et al. \citep{cao2017network} improved community detection in social networks by measuring connection diversity.

\end{itemize}
.

\section{Application of Entropy in science of science}

\subsection{Diversity and Interdisciplinary Research}

\subsubsection{Quick review of Diversity and Interdisciplinary Research}

Diversity and interdisciplinarity are core features of scientific research, closely tied to the growth of scientometrics. As far back as the 1960s, Price's theory of "cumulative advantage" \citep{price1963little} highlighted the uneven nature of scientific collaboration and citation patterns. This work set the stage for analyzing interdisciplinary networks \citep{garfield1964use}. Early scientometric studies, using citation and co-citation analysis \citep{small1973co}, showed that new fields often arise where different disciplines meet, forming what are called "invisible colleges" \citep{crane1972invisible}. In the 1970s, visualizations of journal-to-journal citation networks further confirmed this, uncovering how knowledge moves across disciplines \citep{leydesdorff1986development}.

Since the start of the 21st century, big data tools have fueled quantitative studies of interdisciplinary research. Evidence suggests that papers blending knowledge from various fields tend to make a bigger splash. For instance, Uzzi et al. \citep{uzzi2013atypical} discovered that articles mixing typical ideas with fresh, unconventional ones get cited twice as often as those sticking to standard approaches. Yet, interdisciplinary work faces hurdles. Mainstream databases like Scopus and WoS don’t fully cover fields like social sciences and humanities \citep{lariviere2006place}, which hides some interdisciplinary results. On top of that, evaluation systems can work against it—despite its potential for innovation, interdisciplinary research has a funding success rate about 14\% lower than single-discipline studies \citep{bromham2016interdisciplinary}. Reviewers often miss its originality too \citep{wang2013quantifying}.

Today, team science is reshaping how interdisciplinary research happens. Wuchty et al. \citep{wuchty2007increasing} found that team-written papers jumped from 50\% to 90\% between 1990 and 2000. Large teams pull in resources from hot fields to rack up citations fast \citep{wu2019large}, while smaller teams are more likely to spark breakthroughs by mixing forgotten ideas \citep{uzzi2013atypical}. This split points to a tension in research: systems focused on efficiency, like impact factors, lean toward small, safe steps forward, but big leaps need room for risk \citep{foster2015tradition}. New scientometric tools try to bridge this gap. The SNIP indicator adjusts for field differences \citep{moed2010measuring}, and the I3 index combines productivity and impact to measure interdisciplinary effects \citep{leydesdorff2011integrated}. Still, these methods struggle to fully grasp the value of blending knowledge in new ways.

\subsubsection{Application of Entropy in Diversity and Interdisciplinary Research}

\section*{Citation-based Diversity Analysis}

\begin{itemize}
      \item \textbf{ measuring the diversity of citations in scientific papers and patents:}
Park et al. applied normalized entropy to quantify citation diversity in 45 million papers and 3.9 million patents (1945–2010). Entropy measured the breadth of prior work cited, with higher values indicating engagement across diverse literature. Linked to the CD index (\(CD_t = 1 - \frac{\sum_{i=1}^{n} [2f_{it}b_{it} - f_{it}]}{n_t}\)), entropy correlated positively with disruptiveness. Declining entropy over time mirrored reduced CD scores, revealing narrowing knowledge scope drives decreasing innovation disruptiveness.
\citep{park2023papers}

  \item \textbf{Measuring Interdisciplinary Research Activities through citation flows}:
Wagner et al. applied the concept of information entropy to measure Interdisciplinary Research . They utilized entropy as an indicator of disciplinary impact by analyzing the intensity of knowledge flows between research fields. Specifically, they divided the scientific system into nine fields and calculated entropy values by examining the distribution of citations from other fields to a particular field. This approach provides a forward-looking perspective for assessing knowledge exchange between disciplines, offering a novel quantitative metric for measuring interdisciplinary research activities.\citep{wagner2011approaches}

  \item \textbf{Journal Disciplinary Classification based on citations and references}:
 Rodríguez applied Shannon entropy to journal disciplinary classification research. By calculating the entropy of journal citations and references across different disciplines, combined with internal citation ratios, he proposed an Entropy-Based Disciplinarity Indicator (EBDI). This indicator classifies journals into four categories: knowledge importers, knowledge exporters, disciplinary, and interdisciplinary journals. The research provides a novel quantitative approach to journal classification by using entropy-based methods to better reflect journals' disciplinary characteristics.\citep{rodriguez2017disciplinarity}Bautista-Puig et al. applied this methodology to green and sustainable science and technology journals indexed in Web of Science. Through comparative analysis with Journal Citation Reports (JCR) metrics, they validated the effectiveness of this entropy-based indicator. Their study demonstrated the robustness of the entropy-based approach in characterizing disciplinary patterns within specialized research domains.\citep{bautista2021role}

\item \textbf{Measuring Temporal Imbalance of Knowledge}:
Yang utilized entropy concepts to measure the temporal diversity of knowledge in academic research. He calculated the proportion of references of different ages within each paper, transformed these into probability distributions, and applied the Shannon entropy formula to quantify the imbalance of these distributions, defining it as temporal imbalance. Combined with two other metrics—temporal variety and temporal disparity—he comprehensively calculated temporal diversity. His research revealed that temporal diversity is negatively correlated with citation rates but positively associated with scientific disruptive potential, indicating that citing older literature may promote scientific innovation while reducing short-term impact.\citep{yang2024temporal}
\end{itemize}

\section*{Topic Diversity}
\begin{itemize}
      \item \textbf{revealing  emerging topics in interdisciplinary domains}
Kim et al.  applied information entropy to interdisciplinary scientific research by combining network analysis and BERTopic embedded topic modeling to identify emerging scientific topics. They utilized information entropy as an optimization parameter to identify models with uneven word distribution across topics, ensuring the generation of topic sets with explicit semantic expression. Their findings revealed multiple emerging topics in interdisciplinary domains, with green technologies and health-related technologies emerging as prominent themes across multiple global interdisciplinary science categories.\citep{kim2024identifying}

  \item \textbf{Assessing Topic Specificity}:

Daud et al. applied information entropy as a topic specificity indicator in academic publication evaluation. They measured topic dispersion by calculating the entropy of publication titles and analyzed its correlation with citation counts. Their research demonstrated a negative correlation between topic specificity and citation counts, indicating that publications with lower entropy (more focused topics) typically receive more citations. This finding provides a text-based alternative or complementary metric for ranking academic entities.
\citep{daud2019correlational}

\item \textbf{Measuring topic diversity to predict academic impact}:

Dong et al. applied information entropy to the field of academic impact prediction. They utilized topic modeling (LDA) to model paper titles and abstracts, thereby extracting topic distributions. Topic diversity was measured by calculating the Shannon entropy of a paper's topic distribution. Specifically, higher entropy values indicate greater topic diversity within a paper, which may represent a more comprehensive or in-depth exploration of a subject, potentially resulting in higher citation counts.
\citep{dong2016can}

\item \textbf{Reflect the diversity of styles or themes in creative works}:

Liu et al. applied information entropy to analyze ``hot streaks'' in creative careers. Using deep learning and network science approaches, they constructed high-dimensional representations of creative trajectories for artists, film directors, and scientists, quantifying exploration and exploitation behaviors by calculating entropy values of thematic distributions across their works. The research revealed that hot streaks typically begin with an ``exploration-exploitation'' transition sequence: individuals first explore diverse themes (high entropy) before focusing on specific directions (low entropy). This pattern was consistently observed across all three domains.

\citep{liu2021understanding}

\end{itemize}

\section*{Measurement of Interdisciplinarity}

\begin{itemize}

      \item \textbf{measuring diversity and interdisciplinarity(Diversity index)}:Mutz reimagined Shannon's information entropy concept in science of science for measuring diversity and interdisciplinarity. He decomposed ``diversity'' into three entropy masses: variety, balance, and disparity, combining these components additively rather than multiplicatively. By employing mutual information for disparity measurement and utilizing statistical estimation approaches, he analyzed research output types and funded project data, demonstrating that diversity can be interpreted as the degree of uncertainty, with journal articles exhibiting the strongest balance across research areas.\citep{mutz2022diversity}

  \item \textbf{Assessing Interdisciplinarity in Academic Journals}:
Leydesdorff and Ràfols established that Shannon entropy serves as a robust indicator for assessing interdisciplinarity in academic journals. Their methodological framework involved calculating entropy values for both cited and citing vectors at the journal level, subsequently employing these measurements as quantifiable metrics of interdisciplinary engagement. Comparative analysis within their vector-based indicator system demonstrated that Shannon entropy exhibited superior discriminative performance relative to the Gini coefficient.\citep{leydesdorff2011indicators}

  \item \textbf{Measuring cross-disciplinary understanding between projects}:
Kumar et al. proposed an entropy-based approach to measure and evaluate interdisciplinary understanding. They employed entropic principles to calculate the "Interdisciplinary Factor" (IDF) between projects. Drawing parallels with information theory, where higher entropy indicates greater uncertainty, a higher IDF suggests greater disciplinary divergence and increased complexity in communication between projects. This metric effectively captures both similarities and differences between projects, while also indicating potential communication barriers and opportunities for valuable interdisciplinary research collaborations. The framework provides quantitative insights into the degree of disciplinary integration and potential challenges in cross-disciplinary dialogue.
\citep{kumar2019stronger}

\item \textbf{calculate interdisciplinarity using Shannon entropy as a diversity metric}:
Carusi and Bianchi applied Shannon entropy, Simpson diversity, and the Rao-Stirling index to investigate journal interdisciplinarity. They innovatively utilized singular value decomposition (SVD) of scholar-journal bipartite networks, projecting journals' positions in low-dimensional space onto concept vectors to calculate interdisciplinarity using Shannon entropy as a diversity metric. In experiments within the Italian information and communication technology domain, this methodology effectively identified interdisciplinary journals, particularly capturing those in hidden fields such as optics and antennas, providing more refined interdisciplinarity analysis compared to traditional approaches.\citep{carusi2020look}

\item \textbf{Quantifying authors' interdisciplinarity to predict academic citations}:

Bhat et al. applied Shannon entropy and Jensen-Shannon divergence to the field of academic citation prediction. They quantified authors' interdisciplinarity by calculating the entropy of the distribution of journals in which each author published their papers. They proposed the formula \(\tilde{H} = -\sum_{i=1}^{m} p_i \log p_i\) to measure the diversity of works, where \(p_i\) represents the frequency of engagement in the i-th style or topic, and m is the number of unique styles or topics. Additionally, they employed Jensen-Shannon divergence to measure academic dissimilarity between co-authors. The study constructed classification models using a sample of approximately 8 million scholarly articles. Results indicated a non-linear relationship between citation counts and interdisciplinarity, where moderate levels of interdisciplinarity yielded higher citation rates. Furthermore, these entropy-based metrics demonstrated significant predictive power in a three-class classification problem.\citep{bhat2015citation}
\end{itemize}

\subsection{Knowledge Mapping and Network Analysis}

\subsubsection{Quick Review of Knowledge Mapping and Network Analysis}

The use of knowledge graphs and network analysis in science started in the 1990s. This was a time when the number of scientific papers was growing rapidly. Researchers wanted better ways to measure their impact. They worked with large datasets like Medline, SPIRES, and NCSTRL, which held plenty of information about how scientists collaborated and cited each other’s work. Using these datasets, they built networks to show these relationships and studied their patterns. They examined features like how many connections each point had (degree distribution), whether the network acted like a "small world" where everything links up quickly, and how groups formed within it (community structures). These efforts uncovered trends in how scientists work together and reference each other, such as a few key players having lots of connections while most had few (power-law distribution), top scientists linking up with each other (rich-club phenomenon), and tight-knit clusters in the network \citep{newman2001structure, newman2006finding, redner1998how}.

As complex network theory took off, scientists improved their methods for studying these networks. They began looking at more advanced types, like networks where connections vary in strength (weighted networks), where the direction of links matters (directed networks), or where multiple layers of connections exist (multilayer networks). These approaches gave a clearer picture of how scientific collaboration works and revealed deeper insights. For example, weighted networks helped researchers explore why collaborations often stick to old patterns (social inertia) and how strong or weak ties shape the network \citep{opsahl2008prominence,ramasco2004self,ramasco2006social}. Meanwhile, multilayer networks let them study collaboration and citation links together, showing how these interact \citep{zhou2012quantifying}.

In recent years, researchers have turned to building models to explain how scientific systems change over time. These models often draw on ideas from physics and statistics. They include concepts like popular items gaining more attention (preferential attachment), things losing relevance as they age (aging effects), and some elements being naturally more appealing (node fitness). For instance, Wang and colleagues developed a model using aging effects to track how citations to papers shift and to forecast their long-term influence \citep{wang2013quantifying}. Such models offer fresh ways to understand how science evolves.

Knowledge graphs and network analysis also help predict what’s next for science. Researchers can use them to guess which papers or scientists might gain more influence down the road \citep{newman2009first,mariani2016identification}. Plus, they’ve come up with ranking tools—like PageRank (famous from Google), CiteRank, and AuthorRank—to measure scientists’ impact \citep{liu2005co, yan2011discovering, radicchi2009diffusion}. These tools assist scientists and policymakers in spotting trends and figuring out who stands out in the research world.

\subsubsection{Application of Entropy in Knowledge Mapping and Network Analysis}

\section*{Collaboration Network Analysis}

\begin{itemize}

    \item \textbf{Assessing institutional and departmental diversity in scientific collaboration networks}:
Li et al. applied Shannon entropy to analyze academic collaboration patterns in large language model (LLM) research. The researchers assessed collaboration diversity by calculating entropy values based on authors' institutional and departmental affiliations, and employed difference-in-difference analysis to examine changes before and after ChatGPT's release. Results indicate that interdisciplinary collaboration increased overall but varied across disciplines, with Computer Science and Social Science showing significant enhancements in collaboration diversity, while health-related fields like Medicine exhibited distinct collaboration patterns.\citep{li2024academic}

  \item \textbf{Investigating Collaboration and Knowledge Structuress}:
Miyashita and Sengoku employed graph entropy to investigate collaboration and knowledge structures in interdisciplinary research projects. Through constructing co-authorship and co-word networks, they calculated entropy values based on edge weight distributions, conducting both cross-sectional (across different levels) and longitudinal (temporal evolution) analyses. Their findings revealed correlations between the complexities of collaboration and knowledge structures, with complexity growth moderating in later project phases, reflecting a strategic shift from promoting interdisciplinary research to integrating research outcomes.\citep{miyashita2021scientometrics}

\item \textbf{analyzing researchers' academic mobility trajectories}:
Floriana Gargiulo and Timoteo Carletti applied metric entropy to analyze researchers' academic mobility trajectories. By computing path entropy in real networks and comparing it with a reshuffled null model, they found that the frequency of low-entropy paths (e.g., trajectories oscillating between two nodes) in empirical data was significantly higher than predictions from the random model, while high-entropy paths (e.g., linear diffusive trajectories) were underestimated. This entropy analysis revealed non-random structures in academic mobility, confirming the critical influence of early career positions on subsequent path choices and demonstrating the system's memory effect on initial states.\citep{gargiulo2014driving}

\item \textbf{optimizing team network structures}:
Yves-Alexandre de Montjoye et al. applied information entropy theory to optimize team network structures in collaborative problem-solving. Using a greedy algorithm, they balanced team cohesion and external connectivity by simultaneously maximizing the entropy of repeated small subgraph patterns (network motifs) and minimizing the number of intra-team edges while penalizing missing motifs. Specifically, instrumental tie strength was quantified using the formula \( S_{i,j} = \sum_{t,b} \frac{c_{i,j,t,b}}{N_{t,b}} \), based on physical co-location data. Strong ties were identified via thresholds (e.g., \( E \geq 12 \) for expressive ties indicating best friendships). Results showed that entropy-structured strong ties networks significantly predicted team performance, outperforming technical skills and personality factors. Only the strongest ties exerted significant effects on problem-solving outcomes.\citep{de2014strength}

\item \textbf{Discovering Synergistic Effects in Scientific Collaboration Networks}:
Riahinia et al. employed Shannon entropy analysis to examine collaborative networks within scholarly publications, utilizing entropy-based metrics to quantify and detect synergistic relationships. By applying Shannon's information theory framework, the researchers computed triadic redundancy measures across three key dimensions: author collaborations, international partnerships, and journal interconnections. This methodological approach enabled the identification of nodes and relationships exhibiting synergistic potential. Such latent synergies typically represent unexploited opportunities between distinct nodes in the network, which, when activated, could facilitate the generation of new knowledge and enhance research capabilities.\citep{riahinia2022synergistic}

  \item \textbf{Investigating Collaboration and Knowledge Structuress}:
Miyashita and Sengoku employed graph entropy to investigate collaboration and knowledge structures in interdisciplinary research projects. Through constructing co-authorship and co-word networks, they calculated entropy values based on edge weight distributions, conducting both cross-sectional (across different levels) and longitudinal (temporal evolution) analyses. Their findings revealed correlations between the complexities of collaboration and knowledge structures, with complexity growth moderating in later project phases, reflecting a strategic shift from promoting interdisciplinary research to integrating research outcomes.\citep{miyashita2021scientometrics}

\item \textbf{advisor-advisee relationship identification in collaboration networks}:
Li et al. proposed an advisor-advisee relationship identification method based on the maximum entropy model (MEM) in academic collaboration networks. By extracting features such as publication counts, collaboration ratios, and time differences in first co-authorship, the MEM was applied to learn feature weights and construct a classifier for relationship type classification and temporal boundary estimation. The method avoids the independence assumption of features, enabling comprehensive consideration of interdependent factors. Experimental validation on DBLP datasets demonstrated over 95\% accuracy in relationship identification and a 1.39-year average error in termination time estimation, outperforming state-of-the-art algorithms. This approach provides a robust framework for relationship mining in social networks, with implications for academic collaboration analysis and recommendation systems.\citep{yong2013advisor}

  \item \textbf{Evaluating Academic Influence of Research Teams using Collaboration Network}:
Employing complex network analysis methodology, Liu et al. constructed an author collaboration network based on journal articles indexed in the Web of Science Core Collection from 2013 to 2017, subsequently identifying distinct research teams within the network. The study implemented a comprehensive evaluation framework that integrated bibliometric indicators with social network analysis metrics, utilizing the entropy weight method to determine optimal weighting coefficients for evaluation criteria. Through this systematic approach, the researchers identified 30 prominent research teams and established an academic influence ranking.  \citep{liu2018evaluate}

\end{itemize}

\section*{Citation and Semantic Network Analysis}
\begin{itemize}
 \item \textbf{Detecting Relational Patterns within Citation Networks}:
Dainelli and Saracco employed the maximum entropy principle to develop a Bipartite Configuration Model (BiCM) - a maximum entropy benchmark framework that characterizes term co-occurrence distributions constrained by empirical network properties such as abstract length and term frequency. Their analytical framework demonstrates the operational value of entropy maximization in detecting relational patterns within citation networks, illustrating how this methodology enhances the interpretability of bibliometric analyses by systematically considering complete semantic content in abstracts rather than relying solely on keyword indices or categorical labels.\citep{dainelli2023bibliometric}

  \item \textbf{Analyzing the evolution of semantic networks over time}:
Cunha et al. applied Shannon entropy to time-varying semantic networks of scientific paper titles. They established a methodology for calculating entropy in clique networks along with its maximum and minimum boundary values, analyzing semantic networks formed by paper titles from Nature and Science over a ten-year period using a sliding time window approach. Their results demonstrate that vertex entropy strongly correlates with its maximum values, while edge entropy exhibits greater variability, effectively reflecting temporal changes in vocabulary diversity and thematic connection patterns. This method facilitates tracking the emergence of new ideas or the consolidation of research themes in scientific literature.\citep{cunha2020shannon}
\end{itemize}

\section*{Structural Analysis of Scientific Networks}

\begin{itemize}

\item \textbf{Community Detection in Social Networks}:
Li et al. applied Shannon entropy to community detection in social networks. They quantified network information as entropy values and conceptualized the community detection process as information loss, developing a dynamic programming optimization model. The method achieves community partition by minimizing inter-community information while maximizing intra-community information. When applied to the citation network of Scientometrics journal, it successfully identified 21 research communities with an algorithmic complexity of O(n²), outperforming most existing methods in computational efficiency.\citep{li2015entropy}

  \item \textbf{Assessing Uncertainty in Network Events}:
Entropy is employed as a quantitative metric for assessing system uncertainty in the analysis of Keyword Association Networks for Network Events (KALN). Specifically, this metric quantifies the degree of semantic uncertainty in network events through calculating the distribution entropy of keyword weights. Within this framework, elevated entropy values correspond to heightened systemic uncertainty in the keyword architecture, indicating that the evolutionary trajectory of network events remains less predictable. Conversely, lower entropy measurements reflect increased system determinism, suggesting more defined patterns in event progression.\citep{xuan2015uncertainty}

  \item \textbf{node importance ranking in graph data}:
Liu and Gao proposed a structural entropy-based method for node importance ranking in graph data. The approach leverages information entropy to quantify the structural impact of node removal by calculating the local entropy of connected components based on node degree distributions and integrating edge weights to construct global structural entropy. By evaluating the entropy change after removing each node, the method systematically assesses the node's contribution to maintaining graph connectivity. Experimental results demonstrate that the proposed method outperforms five benchmark methods in terms of monotonicity, robustness, and accuracy across eight real-world datasets.\citep{liu2023structure}

\item \textbf{quantifying node influence in complex networks}:

Qiao et al. proposed an entropy centrality model based on subgraph decomposition and neighbor node entropy to quantify node influence in complex networks. The model calculates local influence entropy on direct neighbors and indirect influence entropy on two-hop neighbors via path-weighted propagation. By integrating these two components, it comprehensively evaluates node importance considering both immediate and extended network effects. Experimental validation on real-world and artificial networks demonstrated that this approach outperforms traditional centrality metrics in identifying critical nodes, highlighting its effectiveness in capturing structural complexity and influence propagation dynamics.\citep{qiao2017identify}

\end{itemize}

\section*{Scientific Trend Prediction and Identification}

\begin{itemize}

\item \textbf{quantifies the degree of change in betweenness centrality distribution across a network(Predicting Scientific Impact)}
The Kullback-Leibler (KL) divergence is employed to compute Centrality Divergence Metric, a metric quantifies the degree of change in betweenness centrality distribution across a network following the publication of a paper. This metric provides an assessment of a publication's structural impact on the underlying knowledge network.

The Centrality Divergence is defined as the Kullback-Leibler divergence between the betweenness centrality distributions in the baseline and updated networks:
\begin{equation}
\text{CKL}(G_{\text{baseline}}, a) = \sum_{i} p_i \log\left(\frac{p_i}{q_i}\right)
\end{equation}
\noindent where:
\begin{itemize}
\item $p_i = C_B(v_i, G_{\text{baseline}})$ represents the betweenness centrality of node $v_i$ in the baseline network
\item $q_i = C_B(v_i, G_{\text{updated}})$ represents the betweenness centrality of node $v_i$ in the updated network after paper publication
\end{itemize}
A larger Kullback-Leibler divergence indicates a more substantial impact of the paper on the network structure, reflecting a higher degree of structural variation introduced by the publication.
\citep{chen2012predictive}

  \item \textbf{Identifying Emerging Breakthrough Topics through networks}:
Xu et al. developed a novel methodological framework that integrates structural entropy analysis with link prediction techniques to identify emerging breakthrough topics in scientific research. The researchers conceptualized scientific knowledge networks as complex adaptive systems and employed structural entropy metrics to quantify their temporal evolutionary patterns. Through systematic analysis of structural entropy dynamics over time, they established a robust approach for identifying potential breakthrough scientific topics. The methodology was further enhanced through the integration of link prediction algorithms to reinforce semantic relationships between topics, significantly improving the predictive accuracy of emerging scientific developments. This dual-component approach demonstrates considerable potential for early detection of transformative research directions.\citep{xu2022methodology}

  \item \textbf{Identifying Emerging Technologies using co-occurrence and co-authorship networks}:
Zhang et al. applied information entropy to identify emerging general-purpose technologies (EGPTs). They constructed a bi-layer network comprising co-occurrence and co-authorship networks, and employed entropy weighting to integrate three centrality measures (degree, closeness, and betweenness centrality) to quantify technologies' fundamentality, speciality, and sociality. Through validation in information science, they successfully identified six potential EGPTs, including content analysis and semantic analysis, demonstrating the methodology's feasibility.
\citep{zhang2021bi}
   \end{itemize}

\subsection{Academic Evaluation}

\subsubsection{Quick Review of Academic Evaluation}

How we evaluate academic work has always been tied to the growth of science itself. Early on, it depended heavily on peer review and the reputation of academic circles, which made it pretty subjective. By the mid-20th century, the flood of scientific papers pushed people to find new methods. Garfield introduced the Science Citation Index (SCI), which measured a paper’s impact by counting its citations \citep{garfield1955citation}. This shift brought data into the picture and kicked off a numbers-based approach to evaluation \citep{wouters1999history}.

That change sparked fields like scientometrics and bibliometrics. These focus on measuring contributions from researchers, journals, and institutions using citation counts, journal impact factors, and similar tools \citep{moed2005citation}. One well-known measure is the h-index, created by Hirsch. It combines how many papers someone has written with how often they’re cited, offering a handy way to gauge a researcher’s influence \citep{hirsch2005index}.

Since the start of the 21st century, ideas from complex systems science have shaken up academic evaluation. Researchers began treating scientific work like a network that shifts over time. They studied how scientists team up, how papers reference each other, and how knowledge moves around \citep{barabasi2002evolution}. For example, Wu and colleagues found that small teams often spark bold, disruptive ideas, while big teams tend to build on what’s already there \citep{wu2019large}.

There’s also a field called the Science of Science (SciSci) that digs into how science operates using big data and computer models. It looks at things like how a paper’s impact changes over time or predicts where a scientist’s career might head \citep{wang2013quantifying}. Research shows scientists often stick to familiar topics because it’s easier to expand on what they know. This habit can hold back risky, groundbreaking ideas \citep{foster2015tradition}.

Today, evaluating academic work comes with challenges and new twists. Traditional measures have flaws. The h-index, for one, doesn’t adjust for differences across fields or account for teamwork \citep{bornmann2007what}. The journal impact factor can be manipulated and doesn’t say much about a single paper’s quality \citep{lariviere2019journal}. In response, the academic community is pushing for smarter ways to use metrics. Initiatives like the San Francisco Declaration on Research Assessment (DORA) and the Leiden Manifesto argue against leaning too hard on one number. They call for blending qualitative reviews with broader measures \citep{hicks2015leiden}. On the tech side, altmetrics track a paper’s reach on social media, in policy documents, or patents, showing its wider impact \citep{priem2010altmetrics}. Machine learning can even predict a paper’s future citations or spot "sleeping beauty" papers—those that start quiet but later take off \citep{ke2015defining}. Meanwhile, the open science movement is nudging evaluation to value things like data sharing and preprints, not just published articles \citep{mckiernan2016how}.

\subsubsection{Application of Entropy in Academic Evaluation}

\section*{Individual Researcher Evaluation}
\begin{itemize}

  \item \textbf{ Evaluating Research Efficiency}:
Gangan Prathap introduced a novel approach by adapting the concept of entropy from thermodynamics to scientometric analysis, establishing a framework for evaluating scientific productivity and efficiency. In this context, entropy serves as a quantitative measure of information loss that occurs during the statistical aggregation of citation data into mean values. The framework posits that lower entropy values, indicating minimal information loss during aggregation, correlate with higher research efficiency. This thermodynamic analogy provides a sophisticated metric for assessing the consistency and impact of a researcher's scholarly output.\citep{prathap2011comments}

\item \textbf{Analyzing researchers' research productivity trajectories}:

Sunahara et al. applied information entropy to analyze research productivity trajectories by using Dynamic Time Warping (DTW) to compute the similarity matrix of standardized publication output sequences for 8,493 scholars. They then employed Uniform Manifold Approximation and Projection (UMAP) for dimensionality reduction to construct a network representation. The Infomap algorithm, a community detection method based on random walks, was used to minimize the information entropy of random walk paths (quantified by Shannon's entropy formula \(H = -\sum p_i \log p_i\), where \(p_i\) denotes the probability of state transitions). This approach identified six universal productivity patterns: constant, U-shaped, decreasing, periodic-like, increasing, and canonical-like curves. Notably, 74\% of researchers exhibited either increasing or canonical-like trajectories, with the latter showing productivity peaks occurring predominantly in mid-career rather than early stages, challenging the traditional early-peak hypothesis.\citep{sunahara2023universal}

\item \textbf{quantifying researchers' strategy shifts }:

Sunahara et al. analyzed career data of 6,028 Brazilian scientists across 14 disciplines using normalized Shannon entropy to quantify strategy shifts between productivity (\( P \)) and journal impact (\( I \)). The entropy formula \( H = -\sum p_i \log p_i \) measured quadrant occupation patterns, where \( p_i \) denotes time spent in productivity/impact sectors. Non-outlier researchers showed peak entropy \( \approx 0.6 \), indicating strategic preference, while outlier researchers (excluding rare \( IP++ \) cases) demonstrated entropy \( \approx 1 \), reflecting strategy neutrality. Disciplinary differences emerged: mathematics approached random strategy adoption, while physics showed stability. Results highlight researchers' aversion to simultaneous productivity/impact adjustments, with early-career high-impact strategies more prevalent.\citep{sunahara2021association}

\item \textbf{Characterizing the Distribution Pattern of Publication Quality}:
Prathap introduced a novel bibliometric evaluation methodology that draws upon thermodynamic concepts. By incorporating terminologies such as 'energy,' 'work,' and 'entropy' from thermodynamics, he developed a framework for quantifying scholarly output. Specifically, the concept of 'entropy' was adapted to characterize the distribution pattern of publication quality across a scientist's body of work. The relationships between energy, work, and entropy were visualized through phase diagrams, providing an intuitive representation of the temporal evolution patterns in scientific productivity. This thermodynamic analogy offers a sophisticated approach to analyzing the dynamics and quality distribution of research output over a researcher's career trajectory.\citep{prathap2011energy}
Building upon Prathap's foundational work, Franceschini and Maisano conducted further investigations into the thermodynamic analogy for bibliometric assessment. While acknowledging the merit of Prathap's innovative conceptual framework, they identified the need for model refinement to achieve greater theoretical consistency.\citep{franceschini2011analogy}

      \item \textbf{ measuring researchers' academic specialization}
Kim et al. examines the gender pay gap among faculty in a U.S. public university system by applying information entropy (\(H = -\sum_{i} p_{i} \log(p_{i})\)) to measure academic specialization. Using confidence scores of 19 root concepts from the OpenAlex database, the authors calculated entropy values based on the distribution of faculty research topics. Lower entropy indicated higher specialization. Results showed a positive correlation between specialization and salary, yet no gender - specific effects were detected, suggesting that performance metrics like H - index and specialization do not account for persistent pay disparities. The entropy - based approach provided a novel method to quantify disciplinary focus and its impact on remuneration in academic settings.\citep{kim2024persistent}

   \item \textbf{Examining the Evolution of Scientific Writing:}
Sun et al. developed a novel methodology combining Kullback-Leibler divergence (KLD) and word embedding-based concreteness/imageability analysis to examine the evolution of scientific writing in Philosophical Transactions of the Royal Society (PTRS) from 1665 to 1869. By calculating KLD for lemmas and POS trigrams between adjacent decades, along with changes in word embedding-based concreteness, they revealed that while scientific writing generally trended toward professionalization and abstraction, this evolutionary trajectory was significantly influenced and interrupted by sociocultural factors.\citep{sun2021evolutionary}

 \item \textbf{Measuring author contributions through entropy maximization}:
Gerchak applied the Shannon entropy maximization principle to infer authors' relative contributions in academic publications. By analyzing deviations from alphabetical ordering in author names, he established a contribution difference threshold as a constraint, under which entropy was maximized to deduce the relative contribution ratios among authors. This method provides a quantitative framework for evaluating individual contributions in multi-authored papers, offering an objective reference for academic assessment and evaluation procedures.
This approach suggests a novel quantitative method for addressing the long-standing challenge of attribution in collaborative academic work, particularly in fields where alphabetical author ordering is the default convention.\citep{gerchak2020inferring}

  \item \textbf{Performance Evaluation of Scientific Researchers}:
Sheng et al. investigated a performance evaluation system for university researchers utilizing an entropy weight-TOPSIS integrated approach. In their study, the entropy weight method was systematically applied to determine objective weights for evaluation indicators, while the TOPSIS (Technique for Order Preference by Similarity to Ideal Solution) comprehensive evaluation method was employed to establish a mathematical model. This model facilitates quantitative analysis of multidimensional research performance metrics, incorporating critical factors such as overall performance allocation, research output typology and quantity, individual contribution scores, disciplinary variations, professional experience (including tenure duration and position rank), talent cultivation outcomes, new project initiation, and award hierarchy recognition.
Through rigorous empirical validation and sensitivity analysis, the researchers demonstrated the model's operational feasibility and methodological effectiveness. Furthermore, they developed a weighted averaging algorithm to enhance the rationality and equity of team selection processes and individual performance-based resource distribution. This integrated framework addresses the complexity of academic performance assessment while maintaining statistical robustness and practical applicability in institutional settings. \citep{shengresearch}

\end{itemize}

\section*{Journal Evaluation Methods}
\begin{itemize}
  \item \textbf{Evaluate the weights of citation analysis indicators and Altmetrics indicators}:
Zhao and Zhu applied information entropy to academic journal evaluation, proposing a three-dimensional assessment model integrating academic influence, communication power, and innovation. They employed the entropy method to analyze the dispersion and randomness of journal evaluation indicators, calculating information entropy values, coefficients of variation, and weights for each indicator, combining these with factor analysis, principal component analysis, and coefficient of variation methods. This approach enables objective assessment of journals' multidimensional performance while avoiding the limitations of single evaluation methods. By integrating diverse evaluation results through fuzzy Borda counting, they effectively mitigated the Matthew effect in journal evaluation.\citep{zhao2023research}

  \item \textbf{Journal Evaluation using entropy-weighted TOPSIS approach}:
Duan et al. conducted a comprehensive evaluation of 20 economics journals by applying the entropy weight method to assess academic quality. Utilizing the entropy-weighted TOPSIS (Technique for Order Preference by Similarity to Ideal Solution) approach, their study systematically compared this methodology with Principal Component Analysis (PCA) and conventional evaluation frameworks. The findings demonstrated that the entropy-weighted TOPSIS method enables a more objective and comprehensive assessment of journal quality, effectively mitigating the influence of subjective biases. This investigation provides a novel methodological framework for academic journal evaluation, offering enhanced reliability in multi-criteria decision-making processes within scholarly communication research.\citep{duan2015journals}

  \item \textbf{Assessing the discourse power of academic journals}:
Wang et al. innovatively applied the Entropy Weight Method to assess discourse power in academic journal evaluation systems. Through systematic integration of heterogeneous multi-source data and employment of comprehensive analytical approaches - including correlation analysis, composite factor analysis, Entropy Weight Method, Technique for Order Preference by Similarity to Ideal Solution (TOPSIS), and two-dimensional four-quadrant mapping - they conducted a multidimensional assessment of medical, comprehensive, and internal medicine journals. The findings demonstrate that the Entropy Weight Method effectively determines indicator weights while mitigating subjective bias, thereby enhancing the objectivity and reliability of evaluation outcomes. This methodology establishes a robust framework for multi-criteria decision-making in scholarly communication analysis, particularly in handling complex evaluation systems with interdependent indicators.\citep{wang2024research}

  \item \textbf{Journal Quality Evaluation using entropy weight method}:
Xu et al. introduced an entropy weight method based on factor analysis into the field of journal evaluation, establishing a three-dimensional assessment framework. This innovative framework systematically integrates multidimensional information encompassing article performance, academic community engagement, and publishing platform characteristics. By employing Manhattan distance to quantify journal positions within a three-dimensional coordinate system, the methodology enables comprehensive evaluation of journal influence. The study demonstrated that this approach effectively mitigates the impacts of outliers and uneven journal distribution patterns. Notably, the proposed method significantly elevated the rankings of high-quality journals that had been systematically underestimated by the Journal Impact Factor (JIF) metric, while maintaining robust discrimination among mainstream publications.\citep{xu2023multidimensional}

  \item \textbf{Evaluate the Discourse Power of Academic Journals}:
Wang applied the entropy weight method to evaluate the discourse power of academic journals. By calculating entropy values, difference coefficients, and indicator weights, combined with factor analysis and TOPSIS methodology, he conducted a comprehensive assessment of discourse influence and discourse leadership in medical journals. The research established a multi-dimensional, multi-indicator integrated evaluation system for journal discourse power, demonstrating the feasibility and reliability of this evaluation approach.
\citep{wang2024research}
\end{itemize}

\section*{Bibliometric and Citation Pattern Analysis}
\begin{itemize}

\item \textbf{quantifying global scientific knowledge inequality  through citation lenses}:
Gomez et al. introduce a "citational lensing" framework to quantify global scientific knowledge inequality by measuring textual similarity using the Kullback-Leibler divergence (KLD). The key equations are:

\begin{equation}
KLD(c_i \| c_j) = \sum c_i \log \frac{c_i}{c_j}
\end{equation}

where \( c_i \) and \( c_j \) represent the national signature vectors of countries \( i \) and \( j \), respectively.

This similarity metric constructs the text network \( L_{\text{text}}^T \), which is then compared to the citation network \( L_{\text{citation}} \) to derive the citation bias network:

\begin{equation}
L_{\text{distortion}} = L_{\text{citation}} - L_{\text{text}}^T
\end{equation}

Here, \( L_{\text{citation}} \) represents directed citation flows between countries, while \( L_{\text{text}}^T \) denotes the transposed text similarity network.

Empirical analysis reveals persistent positive distortion (over-citation) for core countries (e.g., USA, China) and negative distortion (under-citation) for peripheral nations, illustrating systemic inequality in knowledge dissemination. This framework provides a novel entropy-based approach to quantify citation bias and evaluate research impact.
\citep{gomez2022leading}

\item \textbf{analyzing research citation patterns using t-index}:

Singh combined Shannon entropy with annual average h-index to create a new metric called ``t-index'' for analyzing research citation patterns. By calculating and normalizing the entropy values of yearly citation distributions, he measures the consistency and randomness in publication performance of researchers or institutions. Results demonstrate that t-index effectively identifies consistent research productivity across different time periods, overcoming the seniority bias of h-index and providing a fair method for evaluating research impact of both new and established institutions.
\citep{singh2022t}

\item \textbf{Conducting bibliometric distribution analysis using entropy principles}:
Lafouge and Michel applied Shannon's information entropy principle to bibliometric distribution analysis. By examining entropy changes during information construction processes, they investigated bibliometric laws including Lotka's and Zipf's laws. Specifically, they employed the Maximum Entropy Principle (MEP) to develop mathematical models analyzing entropy maximization conditions in author productivity and keyword distributions, demonstrating the intrinsic connections between these distribution patterns and information entropy.\citep{lafouge2001links}

   \item \textbf{Identifying key transitions in journal citation patterns}:

Leydesdorff and de Nooy proposed a novel metric to determine whether critical transitions exist in journal citation patterns. This metric calculates the Kullback-Leibler divergence of 2013 relative to 2012 (KL$_{2013|2012}$) and the Kullback-Leibler divergence of 2012 relative to 2011 (KL$_{2012|2011}$), then adds these two values and subtracts the Kullback-Leibler divergence of 2013 relative to 2011 (KL$_{2013|2011}$). 

If this value is less than zero, it indicates a critical transition, suggesting that the citation pattern of 2012 represents a discontinuity in historical development, and the citation pattern of 2013 cannot be predicted using data from 2011 and 2012. Through this methodology, the authors identified critical transitions in the citation patterns of several journals, concluding that these journals underwent nonlinear transformation processes.\citep{leydesdorff2017can}

 \item \textbf{Evaluating Article Impact Based On Knowledge Flow}:
  Wang et al. proposed a novel methodology employing the entropy weight method for assessing scholarly article impact. The researchers critically examined existing evaluation metrics - including citation-based indicators and citation network topology parameters - noting their inherent limitations in capturing hierarchical citation significance. To address this methodological gap, the entropy weight method was systematically applied to determine optimal weighting coefficients for three key dimensions: knowledge flow intensity, knowledge diffusion capacity, and knowledge transfer capability. This quantitative framework ultimately enables the calculation of a composite citation impact index, establishing a more robust and differentiated assessment paradigm for academic influence evaluation.
\citep{wang2019evaluating}

\end{itemize}

\section*{Institutional and National Academic Performance Evaluation}

\begin{itemize}
\item \textbf{measuring the impact of large-scale research infrastructure on regional knowledge innovation}:
Based on panel data from 283 Chinese cities spanning 2000-2020, Yang et al. constructed a Digital Infrastructure Index (Digital\_inf) using the entropy method. The index was formulated by calculating information entropy weights through secondary indicators (including mobile phone users, internet users, etc.). Their research revealed that the National Supercomputing Centers (NSC) significantly promoted local and peripheral knowledge innovation (measured by average SCI publications per capita) through fundamental effects such as regional R\&D investment (R\&D\_exp), scientific talent (R\&D\_talent), and digital infrastructure (Digital\_inf). Spatial spillover effects were achieved through geographical proximity, collaboration, and digital neighborhood relationships, with particularly pronounced effects in emerging innovative cities such as Shenzhen.\citep{singh2022t}

  \item \textbf{Evaluating Universities' Research Capabilities}:
Bădin et al. employed the entropy weight method to evaluate research performance in Romanian universities. In their study, they initially standardized publication data from the Web of Science database spanning 2006-2010, subsequently calculated indicator weights through entropy value analysis, and ultimately constructed a composite indicator to assess and rank 34 Romanian higher education institutions. \citep{buadin2018reflecting}

  \item \textbf{Assessing National Academic Performance}:
\c{S}erban et al. developed an entropy-weighted methodology to construct a comprehensive evaluation index system for assessing national academic performance. This multidimensional framework integrates weighted indicators including publication outputs (articles and conference proceedings) through entropy-based aggregation. The study establishes an objective weighting mechanism grounded in information entropy theory, where indicator weights are dynamically assigned according to the heterogeneity of value distributions across evaluated entities. Empirical validation revealed robust alignment between the entropy-optimized metric and authoritative international benchmarking systems, thereby establishing the methodological validity of this data-driven approach in achieving both discriminative capacity and measurement precision. \citep{cserban2017appraisal}

\end{itemize}

\section*{Innovation, Policy, and Talent Ecosystem Evaluation}
\begin{itemize}

  \item \textbf{Evaluating the predictive performance of patent analysis models}:
Madani et al. employed information entropy as a metric to assess the predictive capability of models and the predictive value of variables within the domain of patent analysis. They utilized the reduction in uncertainty, quantified by a decrease in entropy, as a primary indicator to evaluate the efficacy of the models. A lower entropy value signifies a stronger predictive ability of the model and a higher predictive value of the variables. This application underscores the utility of entropy in identifying key terms that may influence the trajectory of technological advancement and innovation.\citep{madani2018keyword}

\item \textbf{Assessing Innovation Potential of Patents}:
Zhang et al. applied Shannon entropy to evaluate patent potential in technological innovation. They constructed an entropy-based indicator system, using entropy as a weighting coefficient with the fundamental criterion that "the more common an indicator is, the less weight it would have." This approach overcomes the "moderation" results of traditional methods by identifying patents with distinctive performance in specific indicators. By combining this with collaborative filtering techniques, the study successfully identified patents with technological innovation potential from 28,509 USPTO Chinese patents, demonstrating the feasibility and reliability of the method.\citep{zhang2017entropy}

  \item \textbf{Evaluating Enterprise Innovation Resilience}:
Xia et al. proposed an integrated approach combining the entropy-weighted TOPSIS method with the FGM(1,1) model for evaluating enterprise innovation resilience. In their methodological framework, the entropy weight method was systematically employed to determine indicator weights, enabling the construction of a comprehensive evaluation system incorporating critical dimensions such as tolerance for talent shortages and R\&D security. This hybrid model was subsequently applied to conduct spatiotemporal measurement and forecasting analyses of corporate innovation resilience across 30 Chinese provinces during the period 2016-2020. The study establishes a quantitative assessment framework that effectively captures both current status and evolutionary trends in regional innovation resilience development.
\citep{xia2023research}

  \item \textbf{Evaluating policy efficiency in government-guided investment funds}:
Cui et al. innovatively integrated the Entropy Weight Method with the Analytic Hierarchy Process (AHP) to establish an evaluation framework for assessing policy efficiency in Chinese government-guided investment funds. In their methodology, the researchers conducted a systematic quantitative analysis of 518 policy documents. The Entropy Weight Method was specifically employed to determine objective weights for secondary indicators, particularly innovation performance and economic performance metrics. This quantitative weighting approach was subsequently combined with Grey Relational Analysis to comprehensively evaluate policy implementation effectiveness, thereby enhancing the methodological rigor and objectivity of their policy assessment framework.
\citep{cui2022policy}

  \item \textbf{Evaluating talent ecosystems using entropy weighting method}:
Liang and Ing conducted an innovative study integrating the Analytic Hierarchy Process (AHP) with the entropy weight method for assessing China's skilled talent ecosystem. Their research methodology involved three systematic phases: First, a hierarchical evaluation framework was constructed comprising 5 primary indicators, 14 secondary indicators, and 34 tertiary indicators. Subsequently, the entropy weight method was employed to determine the objective weights of these indicators, which were then synergistically combined with a Hopfield neural network for comprehensive grade evaluation.
\citep{liang2024research}

   \end{itemize}

\section{Conclusion}

Our study shows how entropy can be useful in the Science of Science (SciSci) field. Entropy started in thermodynamics and grew through information theory. Now, it’s applied in many areas, including the analysis of scientific papers. We examined tools like Shannon Entropy, the Entropy Weight Method, the Maximum Entropy Principle, and Structural Entropy. These methods measure variety, uncertainty, and complexity in science. They provide fresh insights into how research functions across different fields and help improve decisions in science policy, team organization, and research assessment.

One major finding is that entropy can gauge the diversity of scientific knowledge. This is especially clear in networks of citations and collaborations. Higher entropy, which reflects greater knowledge variety, often ties to research that’s more impactful and original. This is valuable for studies blending multiple disciplines, where tools like impact factors or the h-index fall short in capturing how knowledge merges. By exploring the diversity of knowledge flow in citation networks, entropy also highlights understudied research areas, potentially inspiring new directions.

The Entropy Weight Method stands out for its ability to assess research performance fairly using data. It reduces personal bias, offering a dependable way to evaluate journals and teams. Structural Entropy sheds light on the complexity of research networks, which aids in planning resources and organizing teams. Meanwhile, the Maximum Entropy Principle helps us analyze patterns in research data and understand authors’ roles in group projects. It provides a straightforward approach to managing uncertainty.

Entropy also plays a role in shaping science policy. It reveals trends in funding, institutional outcomes, and the expansion of scientific fields. When combined with network analysis and machine learning, it enables us to forecast emerging trends in science, enhancing planning efforts.

Beyond that, entropy strengthens research evaluation. Traditional approaches often overlook individual contributions within team efforts. Entropy, however, looks at network structures and how knowledge is shared, leading to a more balanced assessment. This matters a lot in team science, where it better captures collective achievements. It also offers a more accurate picture of a journal’s or team’s influence, overcoming the shortcomings of impact factors or the h-index.

In summary, entropy proves to be a powerful tool in SciSci. It tracks uncertainty, diversity, and complexity—key factors for evaluating research quality, improving team dynamics, and informing policy. Pairing entropy with other techniques can make assessments clearer and more equitable, keeping up with science’s fast evolution. Looking ahead, researchers should explore how entropy applies to innovation, technology, and the social impacts of science to unlock its full possibilities.

\bibliographystyle{plainnat}
\bibliography{main}

\end{document}